\documentclass[useAMS,usenatbib]{mn2e}
\usepackage{amssymb}
\usepackage{graphics}
\usepackage{comment}

\newcommand{\ctbd}[1]{}

\newcommand{\kms}{\ensuremath{\rm km\,s^{-1}}}
\newcommand{\ms}{\ensuremath{\rm m\,s^{-1}}}

\newcommand{\teffstar}{\ensuremath{T_{\rm eff}}}

\newcommand{\sophie}{OHP}

\newcommand{\hatcur}{HAT-P-2}
\newcommand{\hatcurb}{HAT-P-2\lowercase{b}}

\newcommand{\hatcurLCrprstar}{\ensuremath{0.07227\pm0.00061}}		%
\newcommand{\hatcurLCimp}{\ensuremath{0.395^{+0.080}_{-0.123}}}		%
\newcommand{\hatcurLCdur}{\ensuremath{0.1787\pm0.0013}}			%
\newcommand{\hatcurLCingdur}{\ensuremath{0.0141^{+0.0015}_{-0.0012}}}	%
\newcommand{\hatcurLCP}{\ensuremath{5.6334729\pm0.0000061}}		%
\newcommand{\hatcurLCT}{\ensuremath{2,454,387.49375\pm0.00074}}		% +31
\newcommand{\hatcurLCMT}{\ensuremath{54,387.49375\pm0.00074}}		%

\newcommand{\hatcurSMEteff}{\ensuremath{6290\pm60}}			%
\newcommand{\hatcurSMEzfeh}{\ensuremath{+0.14\pm0.08}}			%
\newcommand{\hatcurSMElogg}{\ensuremath{4.16\pm0.03}}			%
\newcommand{\hatcurSMEvsin}{\ensuremath{20.8\pm0.3}}			%

\newcommand{\hatcurYYm}{\ensuremath{1.36\pm0.04}}			% 
\newcommand{\hatcurYYr}{\ensuremath{1.64^{+0.09}_{-0.08}}}		%	 
\newcommand{\hatcurYYlogg}{\ensuremath{4.138\pm0.035}}			%
\newcommand{\hatcurYYlum}{\ensuremath{3.78^{+0.48}_{-0.38}}}		%
\newcommand{\hatcurYYmv}{\ensuremath{3.31\pm0.13}}			%
\newcommand{\hatcurYYage}{\ensuremath{2.6\pm0.5}}			%

\newcommand{\hatcurRVK}{\ensuremath{983.9\pm17.2}}				%
\newcommand{\hatcurRVgammaK}{\ensuremath{316.0\pm6.0}}
\newcommand{\hatcurRVgammaL}{\ensuremath{88.9\pm10.4}}
\newcommand{\hatcurRVgammaS}{\ensuremath{-19860.5\pm10.2}}

\newcommand{\hatcurPPi}{\ensuremath{86^\circ.72^{+1.12}_{-0.87}}}	%
\newcommand{\hatcurPPlogg}{\ensuremath{4.226\pm0.043}}			%
\newcommand{\hatcurPPar}{\ensuremath{8.99^{+0.39}_{-0.41}}}		%
\newcommand{\hatcurPParel}{\ensuremath{0.06878\pm0.00068}}		%
\newcommand{\hatcurPPrho}{\ensuremath{7.29\pm1.12}}			%

\newcommand{\hatcurPPmlong}{\ensuremath{9.09\pm0.24}}			%
\newcommand{\hatcurPPr}{\ensuremath{1.16^{+0.07}_{-0.06}}}		%
\newcommand{\hatcurPPrlong}{\ensuremath{1.157^{+0.073}_{-0.062}}}	%
\newcommand{\hatcurPPmrcorr}{\ensuremath{0.68}}				%
\newcommand{\hatcurPPteff}{\ensuremath{1540\pm30}}			%

\newcommand{\hatcurXdist}{\ensuremath{119\pm8}}				%

\def\sun{\odot}
\def\eqref#1{Eq.~(\ref{#1})}

\title[Refined parameters of HAT-P-2b]{Refined stellar, orbital and 
	planetary parameters of the eccentric HAT-P-2 planetary system}
\author[A. P\'al et al.]{
\newauthor 
Andr\'as~P\'al$^{1,2,3}$\thanks{E-mail: apal@szofi.net},
G\'asp\'ar~\'A.~Bakos$^{1}$\thanks{NSF fellow},
Guillermo Torres$^{1}$, 	% 
Robert~W.~Noyes$^{1}$,  	% 
\newauthor
Debra~A.~Fischer$^{4}$, 	% 
John~A.~Johnson$^{5}$,  	% 
Gregory W. Henry$^{6}$,		% 
R.~Paul~Butler$^{7}$,   	% 
\newauthor
Geoffrey~W.~Marcy$^{8}$, 	% 
Andrew~W.~Howard$^{8}$, 	% 
Brigitta~Sip\H{o}cz$^{1,3}$,	% 
\newauthor
David~W.~Latham$^{1}$ and	% 
Gilbert~A.~Esquerdo$^{1}$\\	% 
$^{1}$ Harvard-Smithsonian Center for Astrophysics,
	60 Garden street, 
	Cambridge, MA, 02138, USA \\
$^{2}$ Konkoly Observatory of the Hungarian Academy of Sciences, 
        Konkoly Thege Mikl\'os \'ut 15-17,
        Budapest, 1121, Hungary \\
$^{3}$ Department of Astronomy, Lor\'and E\"otv\"os University,
        P\'azm\'any P. st. 1/A,
        Budapest, 1117, Hungary \\
$^{4}$ Department of Physics and Astronomy, San Francisco State University, 
	San Francisco, CA, 94132, USA \\
$^{5}$ Institute for Astronomy, University of Hawaii, 
	Honolulu, HI, 96822, USA \\
$^{6}$ Center of Excellence in Information Systems, Tennessee State University,
	Nashville, TN 37209, USA \\
$^{7}$ Department of Terrestrial Magnetism, Carnegie Institute of Washington, 
	Washington DC, 20015, USA \\
$^{8}$ Department of Astronomy, University of California,
	Berkeley, CA, 94720, USA}

\begin{document}

\label{firstpage}

\date{Accepted \dots, received \dots; in original form \dots}

\pagerange{\pageref{firstpage}--\pageref{lastpage}} \pubyear{2007}

\maketitle

%% abstract

\begin{abstract} 
We present refined parameters for the 
extrasolar planetary system HAT-P-2 (also known as HD~147506),
based on new radial velocity and photometric data.
\hatcurb{} is a transiting extrasolar planet that exhibits 
an eccentric orbit.
We present a detailed analysis of the planetary and stellar parameters, yielding
consistent results for the mass and radius of the
star, better constraints on the orbital eccentricity, and refined
planetary parameters. The improved parameters for the host star
are $M_\star=1.36\pm0.04\,M_\sun$ and $R_\star=1.64\pm0.08\,R_\sun$,
while the planet has a mass of $M_{\rm p}=9.09\pm0.24\,M_{\rm Jup}$ 
and radius of $R_{\rm p}=1.16\pm0.08\,R_{\rm Jup}$. 
The refined transit epoch and period for the planet are
$E=\hatcurLCT$\,(BJD) and $P=\hatcurLCP$\,(days), and the 
orbital eccentricity and argument of periastron are 
$e=0.5171\pm0.0033$ and $\omega=185.22^\circ\pm0.95^\circ$.
These orbital elements allow us to 
predict the timings of secondary eclipses with a reasonable accuracy
of $\sim 15$\,minutes.
We also discuss the effects of this significant eccentricity including the
characterization of the asymmetry in the transit light curve. Simple formulae
are presented for the above, and these, in turn, can be used to
constrain the orbital eccentricity using purely photometric data.
These will be particularly useful for very high precision, space-borne
observations of transiting planets.
\end{abstract}

%% EOF abstract
%% keywords


%% EOF keywords

%% EOF titlepage

%%%%%%%%%%%%%%%%%%%%%%%%%%%%%%%%%%%%%%%%%%%%%%%%%%%%%%%%%%%%%%%%%%%%%%%%%%%%%%

%% Introduction

\section{Introduction}
\label{sec:introduction}

At the time of its discovery, \hatcurb{} was the longest period 
and most massive transiting extrasolar planet (TEP), and the only one 
known to exhibit an eccentric orbit \citep{bakos2007a}. In the following
years, other TEPs have also been discovered with significant
orbital eccentricities and long periods: GJ~436b \citep{gillon2007},
HD~17156b \citep{barbieri2007}, XO-3b \citep{johnskrull2008},
and most notably HD~80606 \citep{naef2001,winn2009b}.
See \texttt{http://exoplanet.eu} for an up-to-date 
database for transiting extrasolar planets.

The planetary companion to \hatcur{} (HD~147506) was detected as a transiting object 
during regular operations of the HATNet telescopes \citep[]{bakos2002,bakos2004}
and the Wise HAT telescope \citep[WHAT, located at 
the Wise Observatory, Israel; see][]{shporer2006}. 
Approximately $26,\!000$ individual photometric measurements of good signal-to-noise ratio (SNR) were gathered with
the HATNet 
telescopes at the Fred Lawrence Whipple Observatory (FLWO, Arizona)
and on Mauna Kea (Hawaii), and with the WHAT telescope.
The planetary transit 
was followed up with the FLWO~1.2\,m telescope and its KeplerCam detector.
The planetary properties have been confirmed by radial velocity
measurements and an analysis of the spectral line profiles. 
The lack of bisector span variations rules out the possibility that
the photometric and spectroscopic signatures are due to a blended background
eclipsing binary or a hierarchical system of three stars.

The spin-orbit alignment of the \hatcur{}(b) system was recently
measured by \cite{winn2007} and \cite{loeillet2008}. Both studies
reported an angle $\lambda$ between the projections of the spin and orbital axes consistent with zero, within an uncertainty of 
$\sim10^\circ$. 
These results are particularly interesting because short period
planets are thought to form at much larger distances and to
then migrate inward.  During this process, orbital eccentricity
is tidally damped, yielding an almost circular orbit \citep{dangelo2006}. 
Physical mechanisms such as Kozai interaction between the transiting planet
and an unknown massive companion on an inclined orbit
could result in tight eccentric orbits \citep{fabrycky2007,takeda2008}. 
However, in such a scenario, the spin-orbit alignment as represented by $\lambda$ can be expected to
be significantly larger. For instance, in the case of
XO-3b, the reported alignments are $\lambda=70^\circ\pm15^\circ$
\citep{hebrard2008} 
and $\lambda=37.3^\circ\pm3.7^\circ$ 
\citep[][although there are indications of
systematic observational effects]{winn2009}.
In multiple planetary systems, planet-planet scattering can also yield
eccentric and/or inclined orbits \citep[see e.g.][]{ford2007}.

The physical properties of the host star \hatcur{} have been
controversial,
since different methods for stellar characterization have resulted in
stellar radii between $\sim1.4\,R_{\odot}$ and $\sim1.8\,R_{\odot}$
\citep[see][]{bakos2007a}. 
Moreover, the true distance to the star has been uncertain in previous studies, with the
Hipparcos-based distance being irreconcilable with the luminosity
from stellar evolutionary models.

In this paper we present new photometric and spectroscopic observations
of the planetary system \hatcur{}(b). The new photometric measurements
significantly improve the light curve parameters, and therefore some
of the stellar parameters are more accurately constrained. Our new
radial velocity measurements 
yield significantly smaller
uncertainties for the spectroscopic properties, including the orbital eccentricity, which have an impact also on
the results of the stellar evolution modeling. In \S~\ref{sec:photometric}
we summarize our photometric observations of this system, and in 
\S~\ref{sec:rv} our new
radial velocity measurements. The details of the analysis are discussed
in \S~\ref{sec:analysis}. We summarize our results in \S~\ref{sec:discussion}.

%% EOF introduction

%%%%%%%%%%%%%%%%%%%%%%%%%%%%%%%%%%%%%%%%%%%%%%%%%%%%%%%%%%%%%%%%%%%%%%%%%%%%%%

%% Photometric observations

\section{Photometric observations and reductions}
\label{sec:photometric}

\begin{figure}
\begin{center}
\resizebox{80mm}{!}{\includegraphics{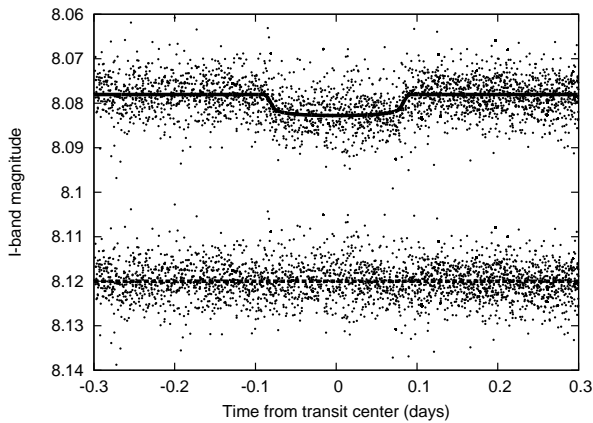}}
\end{center}
\caption{The folded HATNet light curve 
of \hatcur{} \citep[published in][]{bakos2007a}, showing the 
points only near the transit. The upper panel is superimposed with our
best-fit model and the lower panel shows the residuals from the fit.
See text for further details.}\label{fig:hatnetlc}
\end{figure}

\begin{figure}
\begin{center}
\resizebox{86mm}{!}{\includegraphics{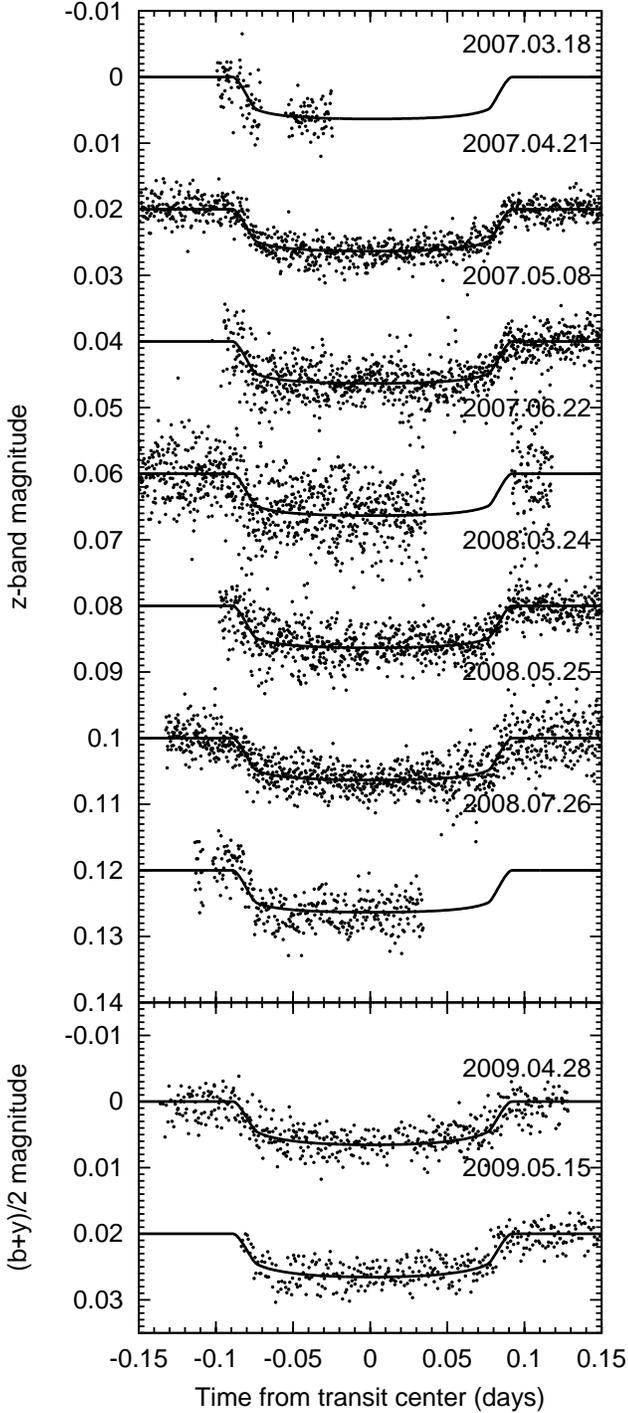}}
\end{center}
\caption{Follow-up 
light curves of \hatcur{}. The top panel shows the $z$-band light curves
acquired on 2007 March 18, 2007 April 21, 2007 May 08, 2007 June 22,
2008 March 24, 2008 May 25 and 2008 July 26; the respective transit 
sequence numbers are $N_{\rm tr}=-6$, $0$, $+3$, $+11$, $+60$, $+71$,
and $+82$. The lower panel shows the Str\"omgren $(b+y)/2$ light 
curves, gathered on 2009 April 28 and 2009 May 15, with
transit sequence numbers $N_{\rm tr}=+131$ and $+134$.
Our best-fit model is superimposed.
See text for further details.}\label{fig:lc}
\end{figure}

\begin{figure}
\begin{center}
\resizebox{80mm}{!}{\includegraphics{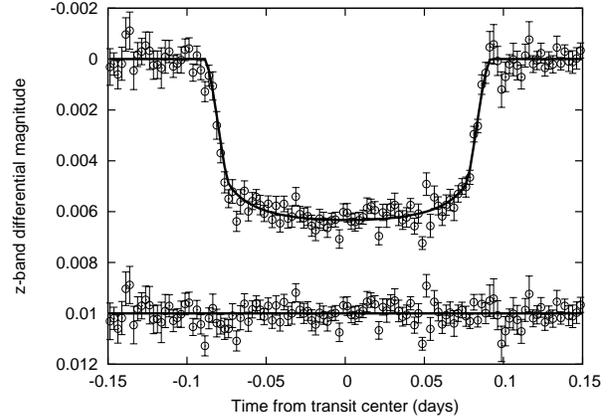}}
\end{center}
\caption{Folded and binned follow-up light curve 
of \hatcur{}, calculated from the seven individual $z$-band events.
The flux values at each point have been derived 
from $\sim$35--50 individual measurements, and the bin size 
corresponds to a cadence of $3.6$ minutes ($0.0025$ days). 
The error bars are derived
from the statistical scatter of the points in each
bin. Typical uncertainties 
are $\sim 0.4$\,mmag.}\label{fig:binlc}
\end{figure}

In the present analysis we make use of photometric data obtained with
a variety of telescope/detector combinations, including the HATNet telescopes, the KeplerCam detector
mounted on the FLWO~1.2\,m telescope, the Nickel 1\,m telescope 
at Lick Observatory on Mount Hamilton, California,
and four of the automated photometric telescopes (APTs) 
at Fairborn Observatory in southern Arizona. 
The photometric analysis of the HATNet data has been described by \cite{bakos2007a}. 
These HATNet data are shown in
Fig.~\ref{fig:hatnetlc}, with our new best-fit model superimposed
(see \S~\ref{sec:analysis} for details on the light curve modeling).
We observed the planetary transit on
nine occasions:  
2007 March 18\footnote{All of the dates are local (MST or HT) calendar 
dates for the first half of the night.} (Sloan $z$ band), 
2007 April 21 ($z$), 
2007 May 08 ($z$), 
2007 June 22 ($z$),
2008 March 24 ($z$), 
2008 May 25 ($z$), 
2008 July 26 ($z$), 
2009 April 28 (Str\"omgren $b+y$ band), and 
2009 May 15 ($b+y$).
These yielded 6 complete or nearly complete transit light curves, and 3 
partial events. One of these follow-up light curves 
(2007 April 21) was reported in the discovery paper. 
All of our individual high precision
follow-up photometry data are plotted in Fig.~\ref{fig:lc}, along with
our best-fit transit light curve model.
The folded and binned light curve (computed only for 
the $z$-band observations) is displayed in Fig.~\ref{fig:binlc}.

The frames taken with the KeplerCam detector were calibrated and 
reduced in the same way for the six nights at FLWO. For the calibrations
we omitted saturated pixels, and applied standard procedures for bias, dark, and sky-flat corrections.

Following the calibration, the detection of stars
and the derivation of the astrometric solution was carried out
in two steps. First, an initial astrometric transformation 
was derived using the $\sim 50$ brightest and non-saturated
stars from each frame, and by using the 2MASS catalogue \citep{skrutskie2006} 
as a reference. We utilized the algorithm of \cite{pal2006}
with a second-order polynomial fit. 
The astrometric data from the 2MASS catalog were obtained from images
with roughly the same SNR as ours. However, we expect significantly
better precision from the FLWO~1.2\,m owing to the larger number 
of individual observations (by two orders of magnitude).
Indeed, an internal catalog
which was derived from the stellar centroids by registering them
to the same reference system has shown an internal precision of
$\sim 0.005$\,arc~sec for the brighter stars, while the 2MASS catalog
reports an uncertainty that is an order of magnitude larger: 
nearly $\sim 0.06$\,arc~sec.
Therefore, in the second step of the astrometry, we used this 
new internal catalog to derive the individual astrometric solutions for
each frame, still using a second-order polynomial fit. We note here
that this method also corrects for systematic errors in the
photometry resulting from the proper motions of the stars, which have changed
their position since the epoch of the 2MASS catalogue ($\sim$2000).

%% %% %% %% %% %% %% %% %% %% %% %% %% %% %% %% %% %% %% %% %% %% %% %% %% %% 
%% %% %% %% %% %% %% %% %% %% %% %% %% %% %% %% %% %% %% %% %% %% %% %% %% %% 

Using the astrometric solutions above we performed aperture photometry
on fixed centroids,
employing a set of five apertures between 7.5 and 17.5 pixels in radius. 
The results of the aperture photometry were then transformed to the same 
instrumental magnitude system using a correction to the spatial distortions and
the differential extinction (the former depends on the celestial 
coordinates while the latter depends on the intrinsic colors 
of the stars). Both corrections were  
linear in the pixel coordinates and linear in the colors. Experience
shows that significant correlations can occur between the
instrumental magnitudes and some of the external parameters
of the light curves (such as the FWHM of the stars, and positions at the sub-pixel level).
Ideally, one should detrend these correlations using only
out-of-transit data (i.e., before ingress and after egress).
Because of the lack of out-of-transit data, we instead carried
out an external parameter decorrelation (EPD) simultaneous
with the light curve modeling (\S~\ref{sec:analysis}) as 
described in \cite{bakos2009}.
After the simultaneous light curve
modelling and de-trending, we chose the aperture for each night that
yielded the smallest residual. 
In all cases this ``best aperture'' was neither the smallest
nor the largest one from the set, confirming the requirement to
select a good aperture series. We note here that since all
of the stars on the frames were well isolated, such choice
of different radii for the apertures does not induce systematics
related to variable blending of stars in different apertures.
In addition, due to the high apparent brightness of \hatcur{} and the comparison 
stars, the frames were acquired under a slightly extrafocal setting (in order to avoid
saturation). This resulted in a different characteristic FWHM for each night.
Thus, the optimal apertures yielding the highest SNR
also have different radii for each night.
Additional and more technical details about the 
photometric reductions
are discussed in Chapter~2 of \cite{pal2009b}.

For the observations at Lick Observatory, we used the Nickel Direct 
Imaging Camera, which is a thinned Loral $2048^2$ CCD with a 
$6.3\arcmin$ square field of view. We observed through a
Gunn $Z$ filter, and used $2\times2$ binning for an effective pixel
scale of $0\farcs37$~pixel$^{-1}$. The exposure times were $25$\,s, with a
readout and refresh time between exposures of $12$\,s. The conditions 
were clear for most of this transit with $\sim1\farcs0$ seeing. 
We defocused the images to draw out the exposure time while 
avoiding saturation for the target and reference stars. We applied the 
flat-field and bias calibrations, and determined the instrumental 
magnitude of \hatcur{} using custom routines written in IDL
as described previously by \cite{winn2007b} and \cite{johnson2008}.
We measured the flux of the target relative to two comparison stars
using an aperture with a $17$-pixel radius and a sky background annulus
extending from $18$ to $60$ pixels.

All four of the APTs at Fairborn Observatory have two
channel photometers that measure the 
Str\"omgren $b$ and Str\"omgren $y$ count rates simultaneously
\citep{henry1999}. Since the Str\"omgren $b$ and $y$ bands are fairly close
together and do not provide any useful color information for such shallow
transits, we averaged the $b$ and $y$ differential magnitudes to create a
$(b+y)/2$ ``band pass'', which gives roughly a $\sqrt{2}$ improvement in
precision. The comparison star for all of the APT observations is HD~145435.

%% EOF Photometric observations

%% Follow-up observations

%% %% %% %% %% %% %% %% %% %% %% %% %% %% %% %% %% %% %% %% %% %% %% %% %% %% 
%% %% %% %% %% %% %% %% %% %% %% %% %% %% %% %% %% %% %% %% %% %% %% %% %% %% 

%% RV observations

\section{Radial velocity observations}
\label{sec:rv}

In the discovery paper for \hatcurb{} \citep{bakos2007a}
we reported thirteen individual radial velocity measurements
from HIRES on the Keck I telescope, and ten radial velocity
measurements from the Hamilton echelle spectrograph at the Lick Observatory \citep{vogt1987}.
In the last year
we have acquired $14$ additional radial velocity measurements
using the HIRES instrument on Keck. In the analysis 
we have incorporated as well the radial velocity data reported by \cite{loeillet2008} obtained with
the OHP/SOPHIE spectrograph. We use only their out-of-transit measurements,
thereby avoiding the measurements affected by the 
Rossiter-McLaughlin effect. 
With these additional $8$ observations, we have a total of $23+14+8=45$ 
high-precision RV data points at hand for a refined analysis.

\begin{figure}
\begin{center}
\resizebox{80mm}{!}{\includegraphics{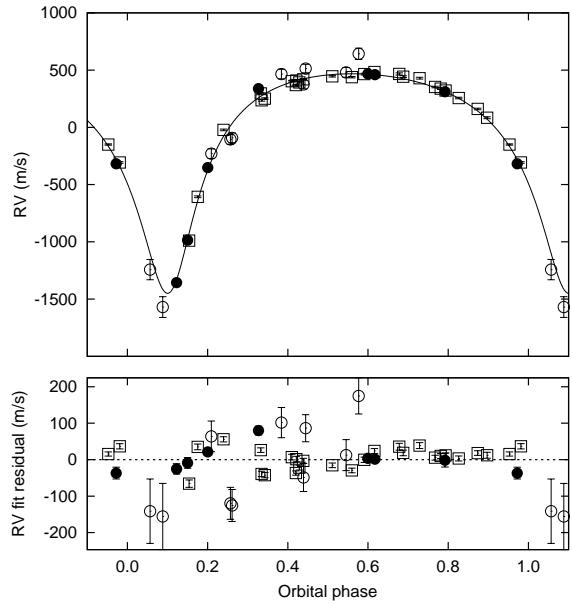}}
\end{center}
\caption{Radial velocity measurements for 
\hatcur{} folded with
the best-fit orbital period. Filled dots represent
the \sophie{} data, open circles show the the Lick/Hamilton, and
the open boxes mark the Keck/HIRES observations. In the upper panel, 
all of these three
RV data sets are shifted to zero mean barycentric velocity.
The RV data are superimposed with our best-fit model. The lower panel
shows the residuals from the best-fit. Note the different vertical
scales in the two panels. The transit occurs at zero
orbital phase. See text for further details.}\label{fig:rv}
\end{figure}

In Table~\ref{tab:rvs} we list all previously published RV measurements as well
as our own new observations.
These data are shown in Fig.~\ref{fig:rv}, along with
our best-fit model described below.

%% EOF RV observations

%% %% %% %% %% %% %% %% %% %% %% %% %% %% %% %% %% %% %% %% %% %% %% %% %% %% 
%% %% %% %% %% %% %% %% %% %% %% %% %% %% %% %% %% %% %% %% %% %% %% %% %% %% 

%% %% %% %% %% %% %% %% %% %% %% %% %% %% %% %% %% %% %% %% %% %% %% %% %% %% 
\begin{table}
\caption{Complete list of relative radial velocity measurements 
for \hatcur{}}\label{tab:rvs}
\begin{center}\begin{tabular}{lrrc}
\hline
\hline
BJD	& RV	& $\sigma_{\rm RV}$ & Observatory  \\
	& \ms	& \ms	\\
\hline
2453981.77748  &  $    12.0 $ &   $     7.3$ & Keck$^{\rm a}$ \\     
2453982.87168  &  $  -288.3 $ &   $     7.9$ & Keck$^{\rm a}$ \\     
2453983.81485  &  $   569.0 $ &   $     7.3$ & Keck$^{\rm a}$ \\     
2454023.69150  &  $   727.3 $ &   $     7.8$ & Keck$^{\rm a}$ \\     
2454186.99824  &  $   721.3 $ &   $     7.7$ & Keck$^{\rm a}$ \\     
2454187.10415  &  $   711.0 $ &   $     6.7$ & Keck$^{\rm a}$ \\     
2454187.15987  &  $   738.1 $ &   $     6.8$ & Keck$^{\rm a}$ \\     
2454188.01687  &  $   783.6 $ &   $     7.1$ & Keck$^{\rm a}$ \\     
2454188.15961  &  $   801.8 $ &   $     6.7$ & Keck$^{\rm a}$ \\     
2454189.01037  &  $   671.0 $ &   $     6.7$ & Keck$^{\rm a}$ \\     
2454189.08890  &  $   656.7 $ &   $     6.8$ & Keck$^{\rm a}$ \\     
2454189.15771  &  $   640.2 $ &   $     6.9$ & Keck$^{\rm a}$ \\     
2454216.95938  &  $   747.7 $ &   $     8.1$ & Keck \\     
2454279.87688  &  $   402.0 $ &   $     8.3$ & Keck \\     
2454285.82384  &  $   168.3 $ &   $     5.7$ & Keck \\     
2454294.87869  &  $   756.8 $ &   $     6.5$ & Keck \\     
2454304.86497  &  $   615.5 $ &   $     6.2$ & Keck \\     
2454305.87010  &  $   764.2 $ &   $     6.3$ & Keck \\     
2454306.86520  &  $   761.4 $ &   $     7.6$ & Keck \\     
2454307.91236  &  $   479.1 $ &   $     6.5$ & Keck \\     
2454335.81260  &  $   574.7 $ &   $     6.8$ & Keck \\     
2454546.09817  &  $  -670.9 $ &   $    10.1$ & Keck \\     
2454547.11569  &  $   554.6 $ &   $     7.4$ & Keck \\     
2454549.05046  &  $   784.8 $ &   $     9.2$ & Keck \\     
2454602.91654  &  $   296.3 $ &   $     7.0$ & Keck \\     
2454603.93210  &  $   688.0 $ &   $     5.9$ & Keck \\     

2454168.96790  &  $  -152.7 $ &   $     42.1$ & Lick$^{\rm a}$ \\     
2454169.95190  &  $   542.4 $ &   $     41.3$ & Lick$^{\rm a}$ \\     
2454170.86190  &  $   556.8 $ &   $     42.6$ & Lick$^{\rm a}$ \\     
2454171.03650  &  $   719.1 $ &   $     49.6$ & Lick$^{\rm a}$ \\     
2454218.80810  &  $ -1165.2 $ &   $     88.3$ & Lick$^{\rm a}$ \\     
2454218.98560  &  $ -1492.6 $ &   $     90.8$ & Lick$^{\rm a}$ \\     
2454219.93730  &  $   -28.2 $ &   $     43.9$ & Lick$^{\rm a}$ \\     
2454219.96000  &  $   -14.8 $ &   $     43.9$ & Lick$^{\rm a}$ \\     
2454220.96410  &  $   451.6 $ &   $     38.4$ & Lick$^{\rm a}$ \\     
2454220.99340  &  $   590.7 $ &   $     37.1$ & Lick$^{\rm a}$ \\     

2454227.50160  &  $-19401.4 $ &   $      8.8$ & \sophie{}$^{\rm b}$ \\     
2454227.60000  &  $-19408.2 $ &   $      6.5$ & \sophie{}$^{\rm b}$ \\     
2454228.58420  &  $-19558.1 $ &   $     18.8$ & \sophie{}$^{\rm b}$ \\     
2454229.59930  &  $-20187.4 $ &   $     16.1$ & \sophie{}$^{\rm b}$ \\     
2454230.44750  &  $-21224.9 $ &   $     14.1$ & \sophie{}$^{\rm b}$ \\     
2454230.60290  &  $-20853.6 $ &   $     14.8$ & \sophie{}$^{\rm b}$ \\     
2454231.59870  &  $-19531.1 $ &   $     12.1$ & \sophie{}$^{\rm b}$ \\     
2454236.51900  &  $-20220.7 $ &   $      5.6$ & \sophie{}$^{\rm b}$ \\
\hline
\hline
\end{tabular}\end{center}
\noindent $^{\rm a}$ Published in \cite{bakos2007a}. 

\noindent $^{\rm b}$ Published in \cite{loeillet2008}.
\end{table}
%% %% %% %% %% %% %% %% %% %% %% %% %% %% %% %% %% %% %% %% %% %% %% %% %% %% 

%%%%%%%%%%%%%%%%%%%%%%%%%%%%%%%%%%%%%%%%%%%%%%%%%%%%%%%%%%%%%%%%%%%%%%%%%%%%%%

%% Analysis

\section{Analysis}
\label{sec:analysis}

In this section we describe the analysis of the available
photometric and radial velocity data in order to determine the
planetary parameters as accurately as possible. 
%% This paragraph is basically about deriving the limb darkening
%% parameters for the transit modeling.
%%
To model transit light curves taken in optical or near-infrared
photometric passbands, we include the effect of the stellar
limb darkening. We have adopted the analytic formulae of \cite{mandel2002} 
to model the flux decrease during transits under the 
assumption of a quadratic limb darkening law.
Since the limb darkening coefficients are functions
of the stellar atmospheric parameters (such as effective temperature $\teffstar$, 
surface gravity $\log g_\star$, and metallicity), 
the light curve analysis is preceded
by an initial derivation of these parameters  
using the iodine-free template spectrum obtained with the 
HIRES instrument on Keck~I. We employed the Spectroscopy Made Easy
software package \citep[SME, see][]{valenti1996}, supported by the atomic line
database of \cite{valenti2005}. This analysis yields the 
$\teffstar$, $\log g_\star$, $\mathrm{[Fe/H]}$, and the projected
rotational velocity $v\sin i$. When all of these are free parameters, 
the initial SME analysis gives
$\log g_\star=4.22\pm0.14$\,(cgs), $\teffstar=6290\pm110$\,K, 
$\mathrm{[Fe/H]}=+0.12\pm0.08$, and $v\sin i=20.8\pm0.2\,\rm{km\,s^{-1}}$.
%% No paragraph break!
The limb darkening coefficients were then derived for the $z'$, $I$, and
$(b+y)/2$ photometric bands by interpolation, using the tables provided by
\cite{claret2000} and \cite{claret2004}.  The initial values for these
coefficients were used in the subsequent global modeling of the data
(\S~\ref{sec:lcrv}), and also in refining the stellar parameters
through a constraint on the mean stellar density (see below). 
% These are not interesting, but kept for reference:
% $\gamma_1^{(z)}=0.1430$,
% $\gamma_2^{(z)}=0.3615$,
% $\gamma_1^{(I)}=0.1765$, and
% $\gamma_2^{(I)}=0.3688$.
A second SME iteration was then performed with a fixed stellar surface
gravity.  The final limb darkening parameters are
$\gamma_1^{(z)}=0.1419$,
$\gamma_2^{(z)}=0.3634$,
$\gamma_1^{(b+y)}=0.4734$,
$\gamma_2^{(b+y)}=0.2928$,
$\gamma_1^{(I)}=0.1752$, and
$\gamma_2^{(I)}=0.3707$.

%% %% %% %% %% %% %% %% %% %% %% %% %% %% %% %% %% %% %% %% %% %% %% %% %% %% 
%% %% %% %% %% %% %% %% %% %% %% %% %% %% %% %% %% %% %% %% %% %% %% %% %% %% 

% ==========================================================================
%%
\subsection{Light curve and radial velocity parameters}
\label{sec:lcrv}

The first step of the analysis is the determination of the 
light curve and radial velocity parameters. 
The parameters can be classified into three groups. 
The light curve parameters that are related to
the \emph{physical} properties of the planetary system are the
transit epoch $E$, the period $P$, the fractional planetary 
radius $p\equiv R_{\rm p}/R_\star$, the impact parameter $b$, and the 
normalized semi-major axis $a/R_\star$. The physical radial velocity parameters
are the RV semi-amplitude $K$, the orbital eccentricity $e$, and
the argument of periastron $\omega$. In the third group there are
parameters that are not related to the physical properties of the
system, but are rather instrument-specific. 
These are the out-of-transit instrumental magnitudes of the 
followup (and HATNet) light curves, and the zero-points $\gamma_{\rm Keck}$,
$\gamma_{\rm Lick}$, and $\gamma_{\rm OHP}$ of the three individual RV data 
sets\footnote{Since a synthetic
stellar spectrum was used as the reference in the reduction of the 
\cite{loeillet2008} data, $\gamma_{\rm OHP}$ is the
actual barycentric radial velocity of the system. 
In the reductions of the Keck and Lick data we used an observed spectrum
as the template, so the zero-points of these two sets are arbitrary
and lack any real physical meaning.}.

To minimize the correlation between the adjusted parameters, we use 
a slightly different parameter set than that listed above. Instead of adjusting the epoch and
period, we fitted the first and last available transit center times, 
$T_{-148}$ and $T_{+134}$.
Here the indices denote the transit event number:
the $N_{\rm tr}\equiv0$ event was defined as
the first complete follow-up light curve taken on 2007 April 21,
the first available transit observation from the HATNet data was
event $N_{\rm tr}\equiv -148$, and the last complete follow-up event
($N_{\rm tr}\equiv+134$) was observed on 2009 May 15.
Note that if we assume the transit events are equally spaced in time, all of the transit centers
available in the HATNet and follow-up photometry are constrained
by these two transit times.
Similarly, instead of the eccentricity $e$ and argument of periastron $\omega$,
we have used as adjustable parameters the Lagrangian orbital elements $k\equiv e\cos\omega$
and $h\equiv e\sin\omega$. These two quantities have the advantage of
being uncorrelated for all practical purposes. Moreover, the radial velocity curve is an analytic
function of $k$ and $h$ even for cases where $e\to 0$ \citep{pal2009a}.
As is well known \citep{winn2007b,pal2008c}, 
the impact parameter $b$ and
$a/R_\star$ are also strongly correlated, especially for 
small values of $p\equiv R_{\rm p}/R_\star$. Therefore, 
following the suggestion by \cite{bakos2007b},
we have chosen the parameters $\zeta/R_\star$ and $b^2$ for fitting instead
of $a/R_\star$ and $b$, where $\zeta/R_\star$ is related to $a/R_\star$ as
\begin{equation}
\frac{\zeta}{R_\star} = \left(\frac{a}{R_\star}\right)\frac{2\pi}{P}\frac{1}{\sqrt{1-b^2}}\frac{\sqrt{1-e^2}}{1+h}.\label{eq:zetaar}
\end{equation}
The quantity $\zeta/R_\star$ is related to the transit duration as
$T_{\rm dur}=2 (\zeta/R_\star)^{-1}$, the duration here being defined 
between the time instants when the center of the planet crosses the limb
of the star inwards and outwards, respectively.

The actual flux decrease caused by the transiting planet 
can be estimated from the projected radial distance 
between the center of the planet and the center of the star 
$d$ (normalized to $R_\star$).
For circular orbits the time dependence of $d$ is trivial 
\citep[see e.g.][]{mandel2002}.
For eccentric orbits, it is necessary to use a precise parametrization of 
$d$ as a function of time. 
As was shown by \cite{pal2008c}, $d$ can be expressed in a second order 
approximation as 
\begin{equation}
d^2=(1-b^2)\left(\frac{\zeta}{R_\star}\right)^2(\Delta t)^2+b^2, \label{zlinear}
\end{equation}
where $\Delta t$ is the time between the actual transit time and the
RV-based transit center. Here the \emph{RV-based transit center} is defined
when the planet reaches its maximal tangential velocity 
during the transit. 
Throughout this paper we give the ephemeris
for the RV-based transit centers and denote these simply by $T_{\rm c}$.
Although the tangential velocity cannot be measured
directly, 
the RV-based transit center is constrained purely by
the radial velocity data, without requiring any prior knowledge of the transit 
geometry\footnote{In other words, predictions can only be made 
for the RV-based transit center in the cases where the planet 
was discovered by a radial velocity survey and initially there are no
further constraints on the geometry of the system, notably its impact 
parameter.}.
For eccentric orbits the impact parameter $b$ is 
related to the orbital inclination $i$ by
\begin{equation}
b=\left(\frac{a}{R_\star}\right)\frac{1-e^2}{1+h}\cos i.
\end{equation}
In order to have a better description of the transit light curve,
we used a higher order expansion in the $d(\Delta t)$ function 
(Eq.~\ref{zlinear}).
For circular orbits, such an expansion is straightforward. To derive
the expansion for elliptical orbits, we employed the method of Lie-integration
which gives the solution of any ordinary differential 
equation (here, the equations for the two-body problem)
in a recursive series for the Taylor expansion with respect to 
the independent variable (here, the time). By substituting the initial
conditions for a body of which spatial coordinates are written as
functions of the orbital elements, using equations~(C1)--(C8)
of \cite{pal2007} one can derive that the normalized projected distance $d$ up to fourth order is:
\begin{eqnarray}
d^2 & = & b^2\left[1-2R\varphi-(Q-R^2)\varphi^2-\frac13QR\varphi^3\right] + \nonumber \\
 & & \left(\frac{\zeta}{R_\star}\right)^2(1-b^2)\Delta t^2\left[1-\frac13Q\varphi^2+\frac12QR\varphi^3\right], \label{d2expand}
\end{eqnarray}
where 
\begin{equation}
Q  =  \left(\frac{1+h}{1-e^2}\right)^3, \label{corrQ}
\end{equation}
and 
\begin{equation}
R  = \frac{1+h}{(1-e^2)^{3/2}}k. \label{corrR}
\end{equation}
Here $n=2\pi/P$ is the mean motion, and $\varphi$ is 
defined as $\varphi=n\Delta t$. For circular orbits, 
$Q=1$ and $R=0$, and for small eccentricities ($e\ll1$), 
$Q\approx 1+3h$ and $R\approx k$.

\subsection{Joint fit}
\label{sec:jointfit}

Given the physical model parametrized above, we performed a 
simultaneous fit of all of the light curve
and radial velocity data. We used \eqref{d2expand}
to model the light curves, where the parameters $Q$ and $R$ were
derived from the actual values of $k$ and $h$, using equations \eqref{corrQ}
and \eqref{corrR}.
To find the best-fit values for the parameters we employed the downhill
simplex algorithm \citep[see][]{press1992} and we used the method
of refitting to synthetic data sets to infer the probability
distribution for the adjusted values. In order to characterize the
effects of red noise properly, the mock datasets in this bootstrap
method were generated by perturbing randomly only the phases in the Fourier
spectrum of the residuals. The final results of the fit
were 
$T_{-148} =2453379.10210\pm0.00121$,
$T_{+134} =2454967.74146\pm0.00093$,
$K=983.9\pm17.2\,\mathrm{m\,s^{-1}}$,
$k=-0.5152\pm0.0036$,
$h=-0.0441\pm0.0084$,
$R_{\rm p}/R_\star\equiv p=0.07227\pm0.00061$,
$b^2=0.156\pm0.074$,
$\zeta/R_\star=12.147\pm0.046\,\mathrm{day}^{-1}$,
$\gamma_{\rm Keck}=\hatcurRVgammaK\,\mathrm{m\,s^{-1}}$,
$\gamma_{\rm Lick}=\hatcurRVgammaL\,\mathrm{m\,s^{-1}}$, and
$\gamma_{\rm OHP} =\hatcurRVgammaS\,\mathrm{m\,s^{-1}}$. 
The uncertainties
of the out-of-transit magnitudes were in the range 6--21$\times10^{-5}\,{\rm mag}$
for the follow-up light curves, and $16\times10^{-5}\,{\rm mag}$ for the 
HATNet data\footnote{Note that these small uncertainties reflect
only the uncertainties of the instrumental magnitudes, and not the 
intrinsic magnitudes in some absolute photometric system.}. 
The fit resulted in a reduced $\chi^2$ value of $0.992$.
As described in the following 
subsection, the resulting distributions of parameters have been used subsequently as inputs
for the stellar evolution modeling.

%=========================================================================
\subsection{Effects of the orbital eccentricity on the transit}

In this section we summarize how the orbital eccentricity
affects the shape of the transit light curve. 
The leading-order correction term in \eqref{d2expand} in $\varphi$, 
$-2b^2R\varphi$, is related
to the time lag between the \emph{photometric} and RV-based transit 
centers \citep[see also][]{kopal1959}.
The photometric transit center, denoted  $T_{\rm c,phot}$ ,
is defined halfway between the instants when the center of 
the planet crosses the limb of the star inward and outward. 
It is easy to show by solving the equation
$d(\varphi)=1$, yielding two solutions ($\varphi_{\rm I}$
and $\varphi_{\rm E}$), that this phase lag is:
\begin{eqnarray}
\Delta\varphi & = & \frac{\varphi_{\rm I}+\varphi_{\rm E}}{2} = \\
& = & -\frac{b^2 R}{\left(\frac{\zeta}{R_\star}\frac{1}{n}\right)^2(1-b^2)-(Q-R^2)b^2} \approx \\
& \approx & -\left(\frac{a}{R_\star}\right)^{-2}\frac{b^2 k}{(1+h)\sqrt{1-e^2}},
\end{eqnarray}
which can result in a time lag of several minutes. For instance,
in the case of \hatcurb{}, 
$T_{\rm c,phot}-T_{\rm c,RV}=n^{-1}\Delta\varphi=1.6\pm0.9$\,minutes. 

In \eqref{d2expand} the third order terms 
in $\varphi$ describe the asymmetry between the slopes of 
the ingress and egress parts of the light curve. 
For other aspects of 
light curve asymmetries, see \cite{loeb2005} and \cite{barnes2007}.
In cases where no
constraints on the orbital eccentricity are available (such as when there are no RV measurements), 
one cannot treat the parameters $R$ and $Q$ as independent
since the photometric transit center 
and $R$ have an exceptionally high correlation. 
However, if we assume a simpler model function, with only third order terms
in $\varphi$ with fitted coefficients present, i.e.~
\begin{eqnarray}
d^2 & = & b^2\left[1-\varphi^2-\frac13C\varphi^3\right] + \nonumber \\
 & & \left(\frac{\zeta}{R_\star}\right)^2(1-b^2)\Delta t^2\left[1-\frac13\varphi^2+\frac12C\varphi^3\right], \label{d2expand2}
\end{eqnarray}
these will yield a non-zero value for the $C$ coefficient for asymmetric light 
curves. In the case of \hatcurb{}, the derived values for $Q$ and $R$ are
$Q=2.204\pm0.074$ and $R=-0.784\pm0.015$ (obtained from the values of $k$ and $h$;
see \S~\ref{sec:jointfit}). Therefore, the coefficient for the third-order
term in $\varphi$ will be $QR=-1.73\pm0.09$. Using \eqref{d2expand2},
for an ``ideal'' light 
curve (with similar parameters of $k$, $h$, $\zeta/R_\star$ and $b^2$ as for
\hatcurb{}), the best fit value for $C$ will be $C=-2.23$, which is close 
to the value of $QR\approx-1.73$. The difference between  the best fit 
value of $C$ and the fiducial value of $QR$ is explained by the fact that in \eqref{d2expand2}
we adjusted the coefficient for the third order term in $\varphi$
that causes the asymmetry in the light curve, and therefore the corrections
in the lower-order terms 
(such as $-2R$, $Q-R^2$, and $Q/3$ in Eq.~\ref{d2expand}) have been neglected.

%AP10: conclusions: 
Although this asymmetry can in principle be 
measured directly (without leading to any degeneracy 
between the fit parameters), in practice one needs extreme photometric precision to
obtain a significant detection for a non-zero $C$ parameter. Assuming
a photometric time series for a single transit of \hatcurb{} 
with $5\,{\rm sec}$ cadence where each 
individual measurement has a photometric error of $0.01$\,mmag(!), 
the uncertainty in $C$ will be $\pm0.47$, equivalent to a 5-$\sigma$
detection of the light curve asymmetry. This detection would be difficult
with ground-based instrumentation. For example, for a 1-$\sigma$ detection one
would need to achieve a photometric precision of $0.05$\,mmag at the same cadence,
assuming purely white noise).
Space missions such as {\it Kepler} \citep{borucki2007} will
be able to detect orbital eccentricity of other planets relying only on transit photometry.

%% %% %% %% %% %% %% %% %% %% %% %% %% %% %% %% %% %% %% %% %% %% %% %% %% %% 
\begin{table}
\caption{Stellar parameters for \hatcur{}.}\label{tab:stellar}
\begin{center}\begin{tabular}{lcl}
\hline
\hline
Parameter		&  Value			& Source	\\
\hline
$T_{\rm eff}$ (K)	&  \hatcurSMEteff		& SME$^{\rm a}$ \\
$[\mathrm{Fe/H}]$	&  \hatcurSMEzfeh		& SME \\
$\log g_\star$ (cgs)    &  \hatcurSMElogg		& SME \\
$v \sin i$ (\kms)	&  \hatcurSMEvsin		& SME \\
$M_\star$ ($M_{\sun}$)  &  \hatcurYYm			& Y$^2$+LC+SME$^{\rm b}$ \\
$R_\star$ ($R_{\sun}$)  &  \hatcurYYr			& Y$^2$+LC+SME \\
$\log g_\star$ (cgs)    &  \hatcurYYlogg		& Y$^2$+LC+SME\\
$L_\star$ ($L_{\sun}$)  &  \hatcurYYlum			& Y$^2$+LC+SME \\
$M_V$ (mag)		&  \hatcurYYmv   		& Y$^2$+LC+SME \\
Age (Gyr) 		&  \hatcurYYage			& Y$^2$+LC+SME \\
Distance (pc)		&  \hatcurXdist			& Y$^2$+LC+SME \\
\hline
\hline
\end{tabular}\end{center}
\noindent $^{\rm a}$ SME = ``Spectroscopy Made Easy'' package for analysis
	of high-resolution spectra by \cite{valenti1996}. See text.

\noindent $^{\rm b}$ Y$^2$+LC+SME = Yonsei-Yale isochrones \citep{yi2001},
	light curve parameters, and SME results.

\end{table}
%% %% %% %% %% %% %% %% %% %% %% %% %% %% %% %% %% %% %% %% %% %% %% %% %% %% 

\subsection{Stellar parameters}
\label{sec:stellarparams}

As pointed out by 
\cite{sozzetti2007}, the ratio $a/R_\star$ is a more effective luminosity
indicator than the spectroscopically determined stellar surface gravity. In the 
cases where the mass of the transiting planet is negligible,
the mean stellar density is 
\begin{equation}
\rho_\star \approx \frac{3\pi}{GP^2}\left(\frac{a}{R_\star}\right)^3. \label{rhovsar}
\end{equation}
The normalized semi-major axis $a/R_\star$ can be obtained 
from the transit light
curve model parameters, the orbital eccentricity, and
the argument of periastron (see \eqref{eq:zetaar}).

Since \hatcurb{} is quite a massive planet, 
($M_{\rm p}/M_\star \sim 0.01$), relation~(\ref{rhovsar}) requires
a small but significant correction, which also depends on observable quantities 
\citep[see][for more details]{pal2008b}. For \hatcurb{} this
correction is not negligible because $M_{\rm p}/M_\star$ is comparable
to the typical relative uncertainties in the light curve parameters.
Following \cite{pal2008a} the density of the star can be written as
\begin{equation}
\rho_\star  =  \rho_0 - \frac{\Sigma_0}{R_\star}, \label{rhostar}
\end{equation}
where both $\rho_0$ and $\Sigma_0$ are observables, namely,
\begin{eqnarray}
\rho_0	 & = & \frac{3\pi}{GP^2}\left(\frac{a}{R_\star}\right)^3, \label{rho0star}\\
\Sigma_0 & = & \frac{3K\sqrt{1-e^2}}{2PG\sin i}\left(\frac{a}{R_\star}\right)^2. 
\end{eqnarray}
In \eqref{rhostar} the only unknown quantity is the radius of the star,
which can be derived using a stellar evolution model, and 
it depends on a luminosity indicator\footnote{In practice this is either the surface
gravity, the density of the star, or the absolute magnitude (if a
parallax is available).}, the
effective temperature $\teffstar$ (obtained from the SME analysis), and
the chemical composition $\mathrm{[Fe/H]}$. Therefore, one
can write
\begin{equation}
R_\star=R_\star(\rho_\star,\teffstar,\mathrm{[Fe/H]}). \label{rstar}
\end{equation}
Since both $\teffstar$ and $\mathrm{[Fe/H]}$ are known, 
we may solve for the two unknowns in \eqref{rhostar} and \eqref{rstar}.
Note that in order to solve \eqref{rstar}, supposing its
parameters are known in advance, one needs to make use of a certain stellar
evolution model. Such models are available only in tabulated form, and therefore
the solution of the equation requires the inversion of the interpolating
function on the tabulated data. Thus, \eqref{rstar} is only a symbolical
notation for the algorithm which provides the solution. Moreover, if
the star is evolved, the isochrones and/or evolutionary tracks for the
stellar models can intersect each other, resulting in an ambiguous solution
(i.e., one no longer has a ``function'', strictly speaking).
For \hatcur{}, however, the solution of \eqref{rstar} is definite since
the host star is a relatively unevolved main sequence star.
To obtain the physical parameters (such as the stellar radius)
we used the evolution
models of \cite{yi2001}, and interpolated the values of $\rho_\star$,
$\teffstar$ and $\mathrm{[Fe/H]}$ using the interpolator provided by 
\cite{demarque2004}. 

The procedure described above has been applied to all of the parameters
in the input set in a complete Monte-Carlo fashion \citep[see also][]{pal2008a}, 
where the values of $\rho_0$ have been derived 
from the values of $a/R_\star$ and the orbital period $P$ using
\eqref{rho0star}, while the values for $\teffstar$ and $\mathrm{[Fe/H]}$
have been drawn from Gaussian distributions with the mean and
standard deviation of the first SME results 
($\teffstar=6290\pm110$\,K and $\mathrm{[Fe/H]}=+0.12\pm0.08$).
This step produced the probability distribution of the 
physical stellar parameters, including the surface gravity. The 
value and associated uncertainty for that particular quantity is $\log g_\star=4.16\pm0.04$\,(cgs),
which is slightly smaller than the result from the SME analysis.
To avoid systematic errors in $\teffstar$ and $\mathrm{[Fe/H]}$ stemming from their correlation with the spectroscopically determined (and usually weakly constrained) $\log g_\star$, 
we repeated the SME analysis by fixing the value of $\log g_\star$
to the above value from the modeling. This second SME run gave $\teffstar=6290\pm60\,{\rm K}$
and $\mathrm{[Fe/H]}=+0.14\pm0.08$. We then updated the values for
the limb darkening parameters,
%(
%$\gamma_1^{(z)}=0.1419$,
%$\gamma_2^{(z)}=0.3634$,
%$\gamma_1^{(b+y)}=0.4734$,
%$\gamma_2^{(b+y)}=0.2928$,
%$\gamma_1^{(I)}=0.1752$, and
%$\gamma_2^{(I)}=0.3707$
%),
and repeated the simultaneous light curve
and radial velocity fit. The results of this fit were then used to
repeat the stellar evolution modeling, which yielded among other
parameters $\log g_\star=\hatcurYYlogg$\,(cgs). The change compared to
the previous iteration is small enough that no further iterations were necessary. Our use here of this classic treatment of error propagation instead of
a Bayesian approach in order to derive the final stellar parameters
is essentially determined by the functionalities of the SME package.
In view of the fact that
the surface gravity from the stellar evolution modeling
(constrained by the photometric and RV data) has a significantly
smaller uncertainty than the value delivered by the SME analysis,
we believe this kind of iterative solution and the method of error estimation
are adequate. The stellar parameters are 
summarized in Table~\ref{tab:stellar}, and the light curve
and radial velocity parameters are listed in the top two blocks
of Table~\ref{tab:parameters}.

\subsection{Planetary parameters}

In the two previous steps of the analysis we determined the light 
curve, radial velocity curve, and stellar parameters. In order to obtain the
planetary parameters, we combined the two Monte-Carlo data sets 
providing probability distributions for all quantities in a consistent way.
For example, the mass of the planet is calculated using
\begin{equation}
M_{\rm p}=\frac{2\pi}{P}\frac{K\sqrt{1-e^2}}{G\sin i}
\left(\frac{a}{R_\star}\right)^2 R_\star^2,
\end{equation}
where the values for the period $P$, RV semi-amplitude $K$, eccentricity
$e$, inclination $i$, and normalized semi-major axis $a/R_\star$
were taken from the results of the light curve and RV fit, while
the values for $R_\star$ were taken from the corresponding
stellar parameter distribution.
From the distribution of the planetary parameters, 
we obtained the mean values and uncertainties. We derived
$M_{\rm p}=\hatcurPPmlong\,M_{\rm Jup}$ for
the planetary mass and $R_{\rm p}=\hatcurPPrlong\,R_{\rm Jup}$
for the radius, with a correlation coefficient 
of $C(M_{\rm p},R_{\rm p})=\hatcurPPmrcorr$ between these two parameters. 
The planetary parameters 
are summarized in the third block
of Table~\ref{tab:parameters}.
Compared with the values reported by \cite{bakos2007a}, the mass
of the planet has not changed significantly 
(from $M_{\rm p}=9.04\pm0.50\,M_{\rm Jup}$), but the 
uncertainty is now smaller by a factor of two. The new estimate
of the planetary radius is larger by roughly $2$-$\sigma$,
while its uncertainty is similar or slightly smaller than before
\citep[$R_{\rm p}=0.982^{+0.038}_{-0.105}\,R_{\rm Jup}$ for][]{bakos2007a}.

The surface temperature of the planet is poorly constrained
because of the lack of knowledge about redistribution
of the incoming stellar flux or the effects of significant orbital
eccentricity.
Assuming complete heat redistribution, the 
surface temperature can be estimated by time averaging the 
incoming flux, which varies as $1/r^2=a^{-2}(1-e\cos E)^{-2}$ due to the 
orbital eccentricity. The time average of $1/r^2$ is
\begin{equation}
\left<\frac{1}{r^2}\right>=\frac{1}{T}\int\limits_{0}^{T}\frac{\mathrm{d}t}{r^2(t)}=
\frac{1}{2\pi}\int\limits_{0}^{2\pi}\frac{\mathrm{d}M}{r^2(M)},
\end{equation}
where $M$ is the mean anomaly of the planet. Since $r=a(1-e\cos E)$
and $\mathrm{d}M=(1-e\cos E)\mathrm{d}E$, where $E$ is the eccentric anomaly,
the above integral can be calculated analytically and the result is 
\begin{equation}
\left<\frac{1}{r^2}\right>=\frac{1}{a^2\sqrt{1-e^2}}. \label{semicorr}
\end{equation}
Using this time-averaged weight for the incoming flux, we 
derived $T_{\rm p}=\hatcurPPteff\,{\rm K}$. However, the planet surface 
temperature would be $\sim 2975\,{\rm K}$ on the dayside during periastron
assuming no heat redistribution, while the equilibrium temperature
would be only $\sim 1190\,{\rm K}$ 
at apastron. Thus, we conclude that the surface
temperature can vary by a factor of $\sim 3$, depending on the
actual atmospheric dynamics.

%% %% %% %% %% %% %% %% %% %% %% %% %% %% %% %% %% %% %% %% %% %% %% %% %% %% 
\begin{table}
\caption{Spectroscopic and light curve solutions 
for \hatcur{}, and inferred planetary parameters.}\label{tab:parameters}
\begin{center}\begin{tabular}{lc}
\hline
\hline
Parameter		&	Value		\\
\hline
Light curve parameters		\\

~~ $P$ (days)			        	& $\hatcurLCP$ 		\\
~~ $E$ ($\mathrm{BJD}-2,\!400,\!000$)		& $\hatcurLCMT$		\\
~~ $T_{14}$ (days)$^{\rm a}$ 			& $\hatcurLCdur$	\\
~~ $T_{12} = T_{34}$ (days)$^{\rm a}$ 		& $\hatcurLCingdur$	\\
~~ $a/R_\star$			               	& $\hatcurPPar$		\\
~~ $R_{\rm p}/R_\star$		              	& $\hatcurLCrprstar$	\\
~~ $b \equiv (a/R_\star)\cos i (1-e^2)/(1+h)$	& $\hatcurLCimp$	\\
~~ $i$ (deg)			               	& $\hatcurPPi$  	\\

Spectroscopic (RV) parameters	\\
~~ $K$ (\ms)			               	& $\hatcurRVK$		\\
~~ $k\equiv e\cos\omega$			& $-0.5152\pm0.0036$	\\
~~ $h\equiv e\sin\omega$			& $-0.0441\pm0.0084$ 	\\
~~ $e$						& $0.5171\pm0.0033$	\\
~~ $\omega$					& $185.22^\circ\pm0.95^\circ$		\\

Planetary parameters		\\
~~ $M_{\rm p}$ ($M_{\rm Jup}$)			& $\hatcurPPmlong$	\\
~~ $R_{\rm p}$ ($R_{\rm Jup}$)			& $\hatcurPPrlong$	\\
~~ $C(M_{\rm p},R_{\rm p})$			& $\hatcurPPmrcorr$	\\
~~ $\rho_{\rm p}$ (g~cm$^{-3}$)			& $\hatcurPPrho$	\\
~~ $a$ (AU)			                & $\hatcurPParel$	\\
~~ $\log g_{\rm p}$ (cgs)			& $\hatcurPPlogg$	\\
~~ $T_{\rm eff}$ (K)				& $\hatcurPPteff$ (see $^{\rm b}$)	\\

Secondary eclipse		\\
~~ $\phi_{\rm sec}$ 				& $0.1868\pm0.0019$		\\
~~ $E_{\rm sec}$ ($\mathrm{BJD}-2,\!400,\!000$)	& $54,388.546\pm0.011$	\\
~~ $T_{14,\rm sec}$ (days)			& $0.1650\pm0.0034$		\\
\hline
\hline
\end{tabular}\end{center}
\noindent $^{\rm a}$ \ensuremath{T_{14}}: total transit duration,
	time between first and last contact; 
	\ensuremath{T_{12}=T_{34}}: ingress/egress time, time between first 
	and second, or third and fourth contact.

\noindent $^{\rm b}$ This effective temperature assumes uniform heat 
	redistribution, while the irradiance is averaged over the entire
	orbit. See text for further details about the issue
	of the planetary surface temperature.
\end{table}
%% %% %% %% %% %% %% %% %% %% %% %% %% %% %% %% %% %% %% %% %% %% %% %% %% %% 

%% %% %% %% %% %% %% %% %% %% %% %% %% %% %% %% %% %% %% %% %% %% %% %% %% %% 
\begin{figure}
\begin{center}
\resizebox{80mm}{!}{\includegraphics{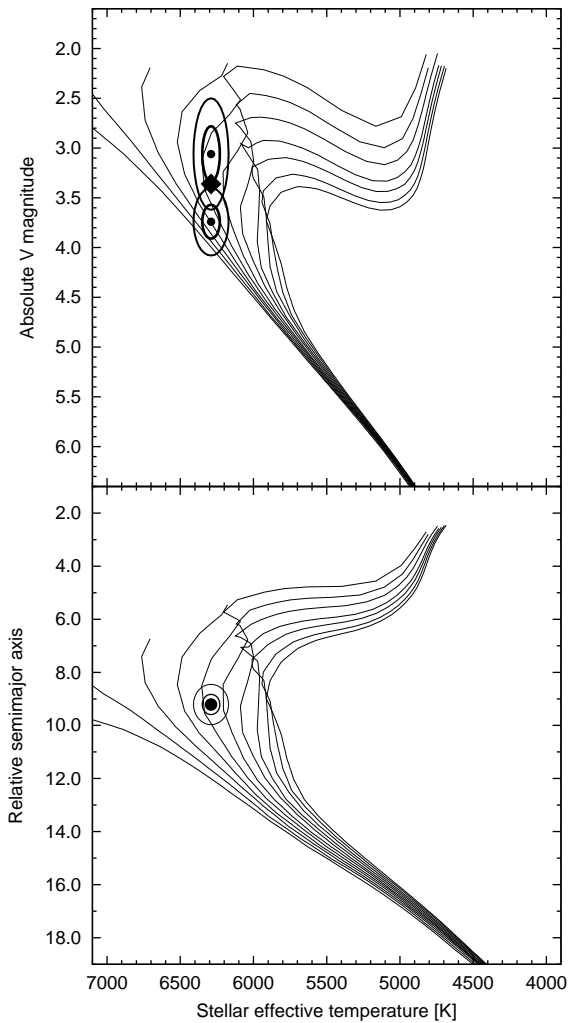}}
\end{center}
\caption{Observational constraints for HAT-P-2 compared with stellar evolution calculations from the Yonsei-Yale models,
represented by isochrones for $[\rm{Fe/H}]=+0.14$ between $0.5$ and 
$5.5$\,Gyr, in steps of $0.5$\,Gyr. The luminosities on the vertical
axis are rendered in two ways: as absolute visual magnitudes $M_{\rm
V}$ in the top panel, and with the ratio $a/R_{\star}$ as a proxy in
the lower panel. The effective temperature along with the absolute
magnitudes inferred from the apparent brightness in the TASS catalog
and the original and revised Hipparcos parallaxes are shown in the top panel with the
corresponding 1-$\sigma$ and 2-$\sigma$ confidence ellipsoids
(upper ellipsoid for the original Hipparcos reductions,
lower for the revision).  The diamond represents the value of $M_{\rm
V}$ derived from our best-fit stellar evolution models using $a/R_{\star}$ as a constraing on the luminosity. In the lower
panel we show the confidence ellipsoids for the temperature and our
estimate of $a/R_{\star}$ from the light curve.\label{fig:isochrones}}
\end{figure}
%% %% %% %% %% %% %% %% %% %% %% %% %% %% %% %% %% %% %% %% %% %% %% %% %% %% 

\subsection{Photometric parameters and the distance of the system}

The measured color index of the star as reported in the 
TASS catalogue \citep{droege2006} is
$(V-I)_{\rm TASS}=0.55\pm0.06$, which is in excellent agreement with the
result of $(V-I)_{\rm YY}=0.552\pm0.016$ we obtain from the stellar evolution 
modeling (see \S~\ref{sec:stellarparams}).
The models also provide the absolute visual magnitude of the star
as $M_V=\hatcurYYmv$, which gives a distance modulus of
$V_{\rm TASS}-M_V=5.39\pm0.13$ corresponding to a distance of
$\hatcurXdist$\,pc, assuming no 
interstellar extinction. This distance estimate
is intermediate between the values inferred from the trigonometric parallax in
the original Hipparcos catalog \citep[$\pi_{\rm HIP} = 7.39\pm0.88$\,mas, corresponding to
a distance of $135\pm18$\,pc;][]{perryman1997}, and in the revised
reduction of the original Hipparcos observations by \cite{vanleeuwen2007a,vanleeuwen2007b} ($\pi_{\rm HIP} = 10.14\pm0.73$\,mas, equivalent to a distance
of $99\pm7$\,pc).
In Fig.~\ref{fig:isochrones} the model isochrones are shown for the 
measured metallicity of \hatcur{} against the measured effective
temperature and two sets of luminosity constraints: those provided
by the estimates of the Hipparcos distance (original, and revised) together with the TASS
apparent magnitudes (top panel), and the constraint from the stellar
density inferred from the light curve (bottom panel).
We note in passing that the distance derived using the near-infrared 2MASS photometry 
agrees well with the distance that relies on the TASS optical magnitudes.

%% %% %% %% %% %% %% %% %% %% %% %% %% %% %% %% %% %% %% %% %% %% %% %% %% %% 
\begin{figure}
\begin{center}
\resizebox{80mm}{!}{\includegraphics{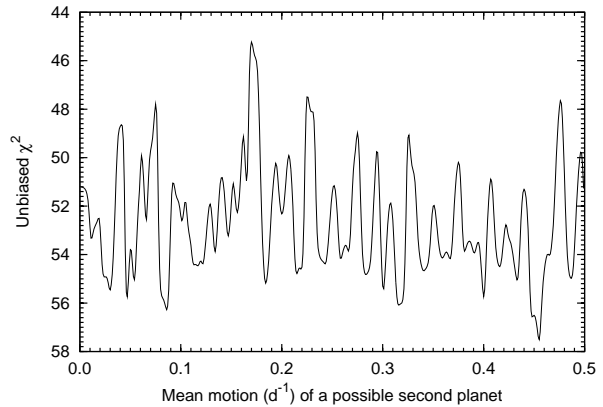}}
\end{center}
\caption{Unbiased $\chi^2$ of a three-body Keplerian + circular
fit to the RV observations, where the mean motion of a possible
secondary companion has been varied between $0$ and $0.5\,{\rm d^{-1}}$.}
\label{fig:rvres}
\end{figure}
%% %% %% %% %% %% %% %% %% %% %% %% %% %% %% %% %% %% %% %% %% %% %% %% %% %% 

\subsection{Limits on the presence of a second companion}

In this section we discuss limits on the presence of an additional
planet in this system.
We performed two types of tests. In both of these
tests we have fitted the RV semi-amplitude $K$, the Lagrangian orbital
elements $(k,h)$, the three velocity zero-points ($\gamma_{\rm Keck}$,
$\gamma_{\rm Lick}$ and $\gamma_{\rm OHP}$), and the additional 
terms required by the respective test methods (drift coefficients
or orbital amplitudes). In these fits, the orbital epoch $E$ and
period $P$ of \hatcurb{} have been kept fixed at the values yielded by the
joint photometric and RV fit. This is a plausible assumption
since without the constraints given by the photometry,
the best fit epoch and period would be 
$E_{\rm RV}=2454342.455\pm0.016$ (BJD) and 
$P=5.6337\pm0.0016$, i.e., the uncertainties would be roughly 20--25
times larger. 

In the first test, a linear, quadratic and cubic polynomial were added to the 
radial velocity model functions in addition to the $\gamma$ zero-point
velocities. Fitting a linear trend yielded a drift of 
$G_{\rm linear}=-21.2 \pm 12.1\,{\rm m\,s^{-1}\,yr^{-1}}$, with
the $\chi^2$ decreasing from $52.1$ to $48.5$
(note that in this test the effective number of degrees of freedom is $45-7=38$).
Therefore, both the decrease in the residuals and the relative 
uncertainty of $G_{\rm linear}$ suggest a noticeable but not very
significant linear drift on the 
timescale of the observations (that is, approximately, $1.7$\,years). 
The additional quadratic and cubic terms do not yield a significant 
decrease in  the unbiased residuals. 

In the second test, we extended the system configuration with an 
additional planet orbiting the star on a circular orbit. The
orbital phase and the semi-amplitude of this additional companion
were fitted simultaneously with the Keplerian orbital elements of \hatcurb{},
while the mean motion of the second companion was varied 
between $n_2=0.001$ and $n_2=0.5\,{\rm d^{-1}}\approx 0.45\,n_{\rm HAT-P-2}$ 
with a step size of $\Delta n=0.001\,{\rm d^{-1}} < (1.7\,{\rm yr})^{-1}$.
As can be seen in Fig.~\ref{fig:rvres},
no significant detection of a possible secondary companion can be 
confirmed. 

\subsection{Secondary eclipse timings}

The improved orbital eccentricity and argument of periastron allow us
to estimate the time of the possible occultations. 
For small orbital eccentricities, the offset of the secondary 
eclipse from phase 0.5 is proportional to $k=e\cos\omega$ \citep[see, e.g.,][]{charbonneau2005}. However, in the 
case of larger eccentricities as in \hatcurb{}, this linear approximation
can no longer be applied. The appropriate formula for arbitrary 
eccentricities can 
be calculated as the difference between the mean orbital longitudes at 
secondary eclipse ($\lambda_{\rm sec}$) and at transit ($\lambda_{\rm pri}$), 
that is,
\begin{eqnarray} 
\lambda_{\rm sec}-\lambda_{\rm pri} & = & 
\pi +\frac{2kJ}{1-h^2} +  \label{eq:phasesec} \\
& + & \arg\left[J^2+\frac{k^2e^2}{(1+J)^2}-\frac{2k^2}{1+J},
2k-\frac{2e^2k}{1+J}\right], \nonumber
\end{eqnarray}
where $J=\sqrt{1-e^2}$. It is easy to see that the expansion of
\eqref{eq:phasesec} yields
\begin{equation}
\lambda_{\rm sec}-\lambda_{\rm pri} \approx \pi + 4k
\end{equation}
for $|k|\ll 1$ and $|h|\ll 1$, and this is equivalent to Eq.~(3) of
\cite{charbonneau2005}.
In the case of \hatcurb{}, we find that secondary eclipses occur at the 
orbital phase of 
$\phi_{\rm sec}=(\lambda_{\rm sec}-\lambda_{\rm pri})/(2\pi)=0.1868\pm0.0019$, 
i.e., 1\,day 1 hour and 17 minutes ($\pm$ 15 minutes) after the
transit events.

%% EOF Analysis

%%%%%%%%%%%%%%%%%%%%%%%%%%%%%%%%%%%%%%%%%%%%%%%%%%%%%%%%%%%%%%%%%%%%%%%%%%%%%%%%

%% Discussion

\section{Discussion}  
\label{sec:discussion}

In this work we have presented refined planetary, stellar and orbital
parameters for the \hatcur{}(b) transiting extrasolar planetary system
based on a full modeling of new and existing data. These data consist
of previously published radial-velocity measurements along with new
spectroscopic observations, and a new set of high-precision
photometric observations of a number of transit events.
The refined parameters have uncertainties 
smaller by a factor of $\sim2$ in the planetary parameters and a factor
of $\sim$3--4 in the orbital parameters than the previously reported 
values of \cite{bakos2007a}. We note that the density
of the planet as determined here, $\rho_{\rm p}=\hatcurPPrho\,{\rm g\,cm^{-3}}$,
is significantly smaller than the value $\rho_{\rm
p,B2007}=11.9^{+4.8}_{-1.6}\,{\rm g\,cm^{-3}}$ inferred by \cite{bakos2007a}, and the new uncertainty is
significantly smaller as well.
Our analysis does not rely on the distance of the system, i.e., we
have not made use of the absolute magnitude as a luminosity
indicator. Instead, our stellar evolution modeling is based on the
density of the star, which is a proxy for luminosity and can be
determined to high precision directly from photometric and RV
observations. A comparison of the distance of the system as derived
from the model absolute magnitude with the Hipparcos determination
(original, and revised) shows that our (density-based) estimate is
intermediate between the two astrometric determinations.

The zero insolation planetary isochrones of \cite{baraffe2003} give
an expected radius of $R_{\rm p,Baraffe03}=1.02\pm0.02\,R_{\rm Jup}$,
which is slightly smaller than the measured radius 
of $\hatcurPPr\,R_{\rm Jup}$. 
The work of \cite{fortney2007} takes into account not 
only the evolutionary age and the total mass of the planet, but the 
incident stellar flux and the mass of the planet's core as well. By scaling
the semi-major axis of \hatcurb{} to one that yields the same incident
flux from a solar-type star on a circular orbit, taking into account
both the luminosity of the star and the correction for the orbital
eccentricity given by \eqref{semicorr}, we derived $a'=0.033\pm0.003\,{\rm AU}$. 
Using
this scaled semimajor axis, the interpolation based on the tables 
provided by \cite{fortney2007} yields radii between 
$R_{\rm p,Fortney,0}=1.142\pm0.003\,R_{\rm Jup}$ (core-less planets) and
$R_{\rm p,Fortney,100}=1.111\pm0.003\,R_{\rm Jup}$ (core-dominated planets, with a
core of $M_{\rm p,core}=100\,M_{\oplus}$). Although these values agree nicely with
our value of $R_{\rm p}=\hatcurPPrlong\,R_{\rm Jup}$, the 
relatively large uncertainty of $R_{\rm p}$ precludes any 
further conclusions about the size of the planet's core. Recent models by
\cite{baraffe2008} also give the radius of the planet as a function of
evolutionary age, metal enrichment, and an optional insolation 
equivalent to $a'=0.045\,{\rm AU}$. Using this 
latter insolation, their models yield 
$R_{\rm p,Baraffe08,0.02}=1.055\pm0.006\,R_{\rm Jup}$ (for metal poor, $Z=0.02$ planets)
and
$R_{\rm p,Baraffe08,0.10}=1.008\pm0.006\,R_{\rm Jup}$ (for more metal rich, $Z=0.10$ planets).
These values are slightly smaller than the actual radius of \hatcurb{}.
However, the actual insolation of \hatcurb{} is roughly two times larger than
the insolation implied by $a'=0.045\,{\rm AU}$. Since the 
planetary radius from \cite{baraffe2008} for zero insolation
gives $R^{(0)}_{\rm p,Baraffe08,0.02}=1.009\pm0.006\,R_{\rm Jup}$
and $R^{(0)}_{\rm p,Baraffe08,0.10}=0.975\pm0.006\,R_{\rm Jup}$ for
metal enrichments of $Z=0.02$ and $Z=0.10$, respectively, an
extrapolation for a two times larger insolation would put the expected planetary
radius in the range of $\sim 1.10\,R_{\rm Jup}$. This is consistent with
the models of \cite{fortney2007} as well as with the measurements. However, 
as discussed earlier in the case of the \cite{fortney2007} models,
the uncertainty in $R_{\rm p}$ does not allow us to properly constrain the metal 
enrichment for the recent Baraffe models.

%\hatcurb{} will remain an interesting target, as a member of an
%emerging heavy-mass population. Further photometric measurements will
%refine the light curve parameters and therefore more precise stellar
%parameters can also be obtained. This will yield smaller uncertainties
%in the physical planetary radius, thus some parameters of the 
%planetary evolution models, such as the metal enrichment can be obtained
%more explicitly. Moreover, observations of secondary eclipses will reveal
%the planetary atmosphere temperature which now is poorly constrained. Since 
%the secondary eclipse occurs shortly after periastron passage, the 
%temperature and therefore the contrast might be high enough to detect
%the occultation with a good signal-to-noise ratio. Moreover, 
%as careful analysis of the RV data series has suggested a possible long-term
%drift, collecting further data points may confirm the presence
%of an additional, long-period companion orbiting the host star.

%% EOF Discussion

%%%%%%%%%%%%%%%%%%%%%%%%%%%%%%%%%%%%%%%%%%%%%%%%%%%%%%%%%%%%%%%%%%%%%%%%%%%%%%%%

%% Acknowledgements

\section*{Acknowledgments}

The work by A.P was supported by the HATNet project
and in part by ESA grant PECS~98073.
HATNet operations have been funded by NASA grants NNG04GN74G, NNX08AF23G 
and SAO IR\&D grants. 
Work of G.\'A.B.~and J.J.\ were supported by the Postdoctoral Fellowship 
of the NSF Astronomy and Astrophysics Program 
(AST-0702843 and AST-0702821, respectively). G.T.\ received
partial support from NASA Origins grant NNX09AF59G.
We acknowledge partial support also from the Kepler Mission under 
NASA Cooperative Agreement NCC2-1390 (D.W.L., PI). 
This research has made use of Keck telescope time granted through NOAO and NASA. 
We thank the UCO/Lick technical staff for supporting the 
remote-observing capability of the Nickel Telescope, allowing the 
photometry to be carried out from UC Berkeley.
Automated Astronomy at Tennessee State University has been 
supported long-term by NASA and NSF as well as Tennessee State 
University and the State of Tennessee through its Centers of Excellence program.
We are grateful for the comments and suggestions by the referee, Frederic
Pont. We acknowledge the use of the VizieR service \citep{ochsenbein2000} 
operated at CDS, Strasbourg, France, of NASA's Astrophysics Data 
System Abstract Service, and of the 2MASS Catalog.

%% EOF Acknowledgements

%%%%%%%%%%%%%%%%%%%%%%%%%%%%%%%%%%%%%%%%%%%%%%%%%%%%%%%%%%%%%%%%%%%%%%%%%%%%%%%%

%% Bibliography

{}

%% EOF Bibliography

\label{lastpage}

\end{document}